# OCTOCAM:
# A fast multichannel imager and spectrograph for the 10.4m GTC

Antonio de Ugarte Postigo*[a], Javier Gorosabel[b], Paolo Spanò[a], Marco Riva[a], Ovidio Rabaza[b], Vincenzo de Caprio[c], Ronan Cuniffe[b], Petr Kubánek[b,d], Alberto Riva[e], Martin Jelínek[b], Michael I. Andersen[f], Alberto J. Castro-Tirado[b], Filippo M. Zerbi[a,g], Alberto Fernández-Soto[h]

[a] INAF-Osservatorio Astronomico di Brera, via E. Bianchi 46, Merate (LC), Italy;
[b] Instituto de Astrofísica de Andalucía (CSIC), Rotonda de la Astronomía, 18008 Granada, Spain;
[c] INAF-IASF, Via E. Bassini, 15 - 20133 Milano Italy;
[d] Image Processing Laboratory, University of Valencia, Valencia, Spain;
[e] Oss. Astronomico di Torino (INAF-OaTo), Str. Osservatorio, 20, 10025 Pino Torinese (TO), Italy;
[f] DARK Cosmol. Centre, Niels Bohr Institute, Juliane Maries vej 30, Copenhagen Ø, Denmark;
[g] ESO, Karl-Schwarzschild-Str. 2, 85748 Garching bei München, Germany;
[h] Instituto de Física de Cantabria (IFCA-CSIC). Edificio Juan Jordá, Av. De los Castros s/n, E-39005, Santander, Spain.

**ABSTRACT**

OCTOCAM is a multi-channel imager and spectrograph that has been proposed for the 10.4m GTC telescope. It will use dichroics to split the incoming light to produce simultaneous observations in 8 different bands, ranging from the ultraviolet to the near-infrared. The imaging mode will have a field of view of 2' x 2' in $u$, $g$, $r$, $i$, $z$, $J$, $H$ and $K_S$ bands, whereas the long-slit spectroscopic mode will cover the complete range from 4,000 to 23,000 Å with a resolution of 700 - 1,700 (depending on the arm and slit width). An additional mode, using an image slicer, will deliver a spectral resolution of over 3,000. As a further feature, it will use state of the art detectors to reach high readout speeds of the order of tens of milliseconds. In this way, OCTOCAM will be occupying a region of the time resolution - spectral resolution - spectral coverage diagram that is not covered by a single instrument in any other observatory, with an exceptional sensitivity.

## 1. INTRODUCTION

OCTOCAM has been thought as a workhorse instrument that optimizes the use of GTC [1][2] for broad-band single-target observations both in imaging and spectroscopy for a wide range of science cases. It is based on the use of high-efficiency dichroics to divide the light into eight different arms, one ultraviolet, four optical and three near-infrared. As compared with traditional instruments, it has the advantage of simultaneity, as it is capable of observing at the same time in $u$, $g$, $r$, $i$, $z$, $J$, $H$ and $K_S$ bands, while keeping a field of view of 2'x2', which is enough for most science cases. With its spectroscopic mode it is capable of obtaining a full spectrum from 4,000 to 23,000 Angstroms in a single shot. Using standard long slit observations, OCTOCAM will provide spectral resolutions that will vary from 700 to 1,700, depending on the selected slit width and arm. Using an image slicer that divides an area of 5"x1.5" into 5 slices of 5"x0.3", it will reach resolutions of over 3000. Furthermore, thanks to the use of state of the art detectors, it will be able to reach high readout speeds, covering science cases aimed at high time-resolution. Depending on the detector of each arm, we expect full frame rates of 17 to 76 Hz, which will be even higher for windowed modes. This will also mean that OCTOCAM will virtually eliminate dead times in most observing modes, allowing duty cycles of almost 100%. Table 1 displays the general specifications of the instrument.

*adeugartepostigo@gmail.com; phone +39 039 5971059; fax +39 039 5971001

Table 1 - OCTOCAM specifications.

| OCTOCAM Specifications | |
|---|---|
| **Simultaneous spectral range** | 0.33 to 2.3 μm (*u, g, r, i, z, J, H* and $K_S$) in imaging<br>0.4 to 2.3 μm in spectroscopy |
| **Field of view** | 2'×2' |
| **Plate scale** | 0.12"/pixel |
| **Detectors** | 1×1k×1k EMCCD (*u* arm)<br>4×2k×2k Frame transfer CCD (*g, r, i* & *z* arms)<br>3×2k×2k HAWAII-2RG (*J, H* & $K_S$ arms) |
| **Average efficiency** | ~ 45 % imaging, ~ 35% spectra |
| **Spectral resolution** | 700-1700 (depending on slit width and arm)<br>~3000 with image slicer |
| **Maximum full-frame rate** | *u* ~ 17Hz, *griz* ~ 40Hz, $JHK_S$ ~ 76Hz |
| **Focus** | Folded Cassegrain |
| **Observing modes** | High-time-resolution imaging<br>Standard imaging<br>High-time-resolution spectroscopy<br>Standard spectroscopy<br>Image slicer spectroscopy (5"×1.5" in 5 slices) |

## 2. INSTRUMENT DESCRIPTION

OCTOCAM is divided into 3 subsystems: the ultraviolet arm (UV), the visible arms (VIS) and the near-infrared arms (NIR) as schematically shown in Figure 1. The first dichroic divides the incoming beam, letting through the VIS and NIR light, while reflecting the light below 4,000 Angstroms towards the UV arm, which has only imaging capabilities and no collimated beam. The rest of the light continues into the cryostat through the first window. The light is focused inside the cryostat, where a cold slit wheel can select one of the various available slits, an image slicer, or allow free path for imaging purposes. Behind the focal plane, the light is again divided into NIR and VIS. The NIR light is collimated by a set of mirrors and then divided into the 3 channels (*J, H, $K_S$*) by further dichroics. The light goes through the filters or grisms (depending on the observing mode) and is then focused by three camera lens systems of equal design. After being reflected by the second dichroic, the VIS light exits the cryostat through the second window. The light is collimated through several lenses and divided into the four VIS arms (*g, r, i, z*) by further dichroics. The light goes through the filters or grisms and is then refocused on the detectors by each of the camera systems. In Figure 1, the *u* and *g* arms are drawn in greater detail to show that the *u*-band has only focal reducer optics and no collimator while the *g* has collimator, grism/filter and camera. The rest of the ams (*r, i, z, J, H, $K_S$*) are equivalent to *g*.

Some of the most important constraints governing the instrument design were the size and weight, as the folded Cassegrain focus accepts a maximum of 1,000 kg. In order to achieve this goal the optics were kept as small as possible and the optical path was folded using mirrors in order to reduce the size of the mechanical elements, and in particular the size of the cryogenic vessel. The resulting instrument has a total combined weight of 980kg (including contingencies), just below the limit of the folded Cassegrain focus of GTC.

In the following subsections we give further details on the characteristics of the design.

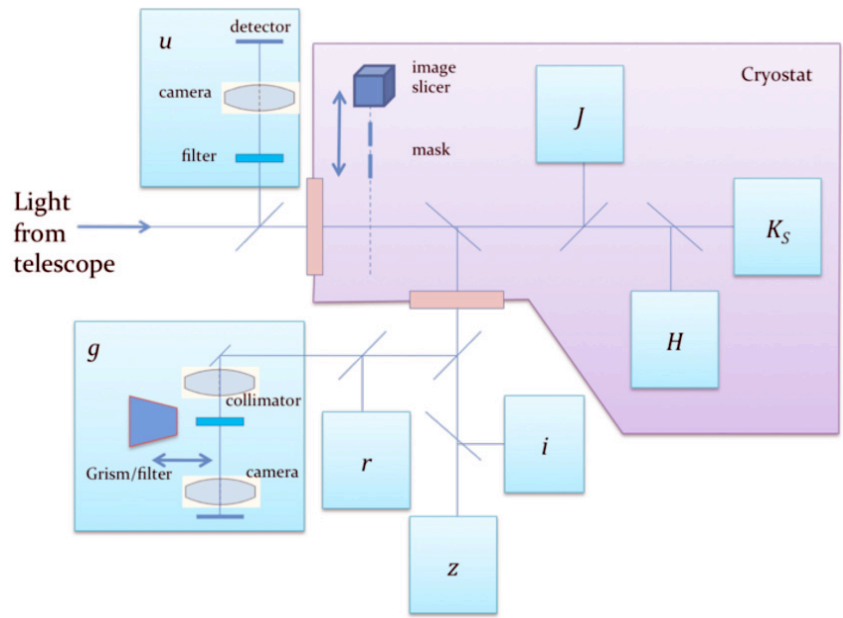

Figure 1 - Schematic view of the instrument.

## 2.1 Optical design

In this section we give details on the preliminary optical design that is foreseen for each of the subsystems of OCTOCAM .

### 2.1.1 UV

The UV arm has a simple design that allows high efficiency by using only UV-transmitting materials, like CaF2 and Silica. No collimated beams are created within the optical path, because the UV arm will be used only in imaging mode and placed before the telescope focal plane. Folding was necessary to fit the UV arm within the given volume (see Figure 2).

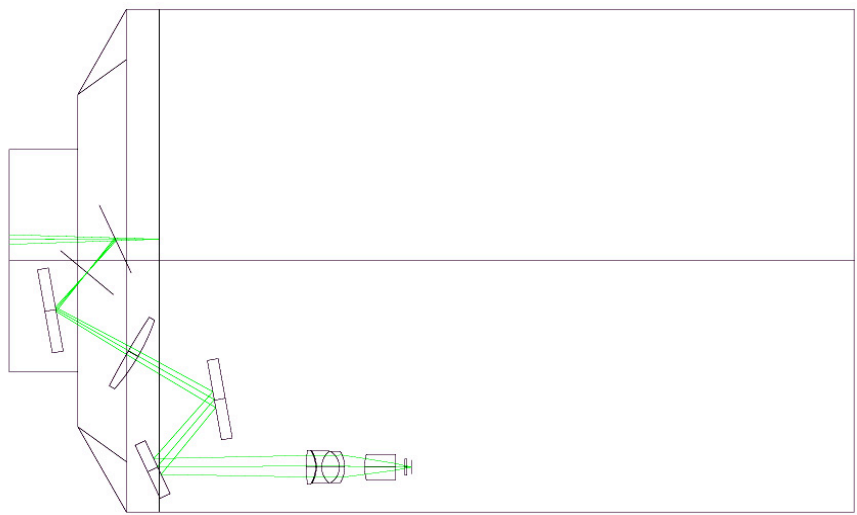

Figure 2 – Optical design of the UV arm within the allocated space for the instrument.

## 2.1.2 VIS

The VIS arm consists of four channels. After the second dichroic, placed inside the cryostat, light from 4,000 to 10,000 Angstroms exits the cryogenic vessel through a flat window. The incidence angle is around 30 degrees to make the design of the cryostat simpler. Collimator optics are based on three doublets, with spherical surfaces and Schott or Ohara glasses selected from those ones available in relatively large glass blanks with good homogeneity. The first element is about 220 mm in diameter, while the other two doublets are <150 mm. A dichroic has been placed after the first element of the collimator optics, working in near collimated beams. Then, folding mirrors have been used to fit the overall instrument layout in the given volume. The cameras are based on 7 lenses split into 4 groups. One aspherical surface has been added to control image quality. All lens diameters are smaller than 80 mm. VPH grisms can be inserted after dichroics and folding mirrors, just in front to the camera lenses. Cameras are located by pairs in two planes, making VPH grisms mountable by pairs on common grism wheels (see Figure 3).

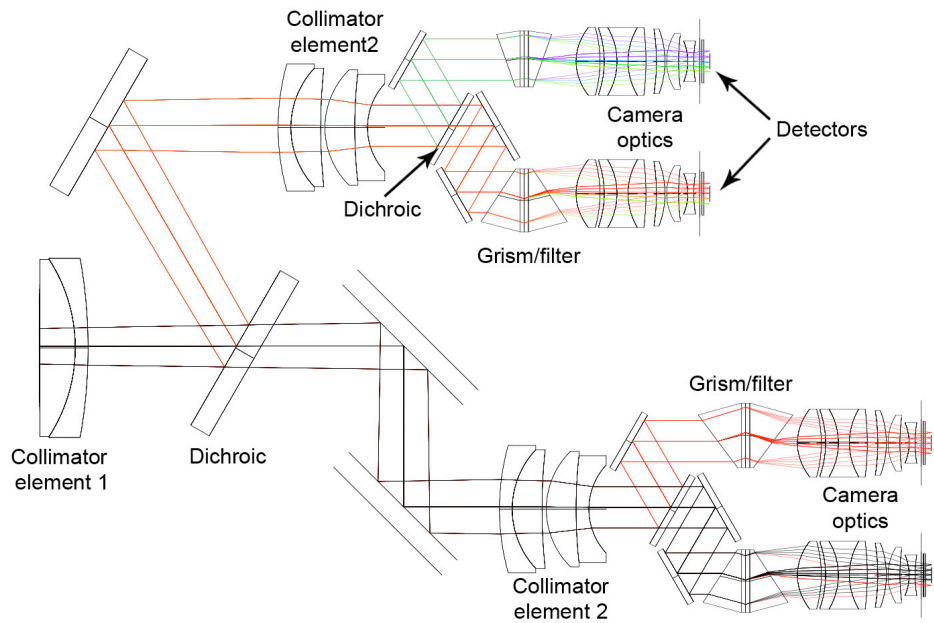

Figure 3 – Optical design of the VIS arms.

## 2.1.3 NIR

A three-mirror off-axis collimator creates a 50 mm pupil image of the telescope entrance. Two mirrors are off-axis portions of aspheres, while the convex second mirror is a spherical one, making it simpler to be manufactured and aligned. The layout is compact to help reduce size and weight of the optics and the related vacuum cryogenic chamber (see Figure 4). Collimator mirrors can be gold coated to enhance efficiency. The largest optics are 200 mm in diameter. A flat folding mirror reflects the beam towards two other dichroics working in a collimated beam, where light is split into three different channels. Dichroics are used at a relatively low incidence angle of 30 degrees. Their size is about 90 mm. The pupil image is focused onto the surface of the last dichroic, to allow a cold mask to be placed before the $K_S$-band channel, which helps to control stray light.

With the help of four flat folding mirrors, light from the dichroics is delivered to three identical cameras, placed at 120 degrees with respect to each other. Such a choice allows a simple rotating wheel as exchange mechanism for the filter/grism unit. Moreover all NIR detectors lie near each other, making cooling simpler and improving stability.

The F/2.3 cameras are based on 6 lenses, with only spherical surfaces, made in standard infrared materials (CaF2, S-FTM16 and Silica), delivering good image quality with high efficiency (>82%). Lens diameters are smaller than 100 mm, for an overall weight of 2 Kg. The last mirror, feeding the $K_S$-band camera, will have tip-tilt capability in order to perform dithering. This allows to correct for the relatively fast varying sky contribution in the $K_S$-band while obtaining longer exposures in the other bands, without the need to move the telescope. Thanks to the compact design, all VPH grisms and filters can be mounted on a single motorized wheel, simplifying the overall design and reducing weight.

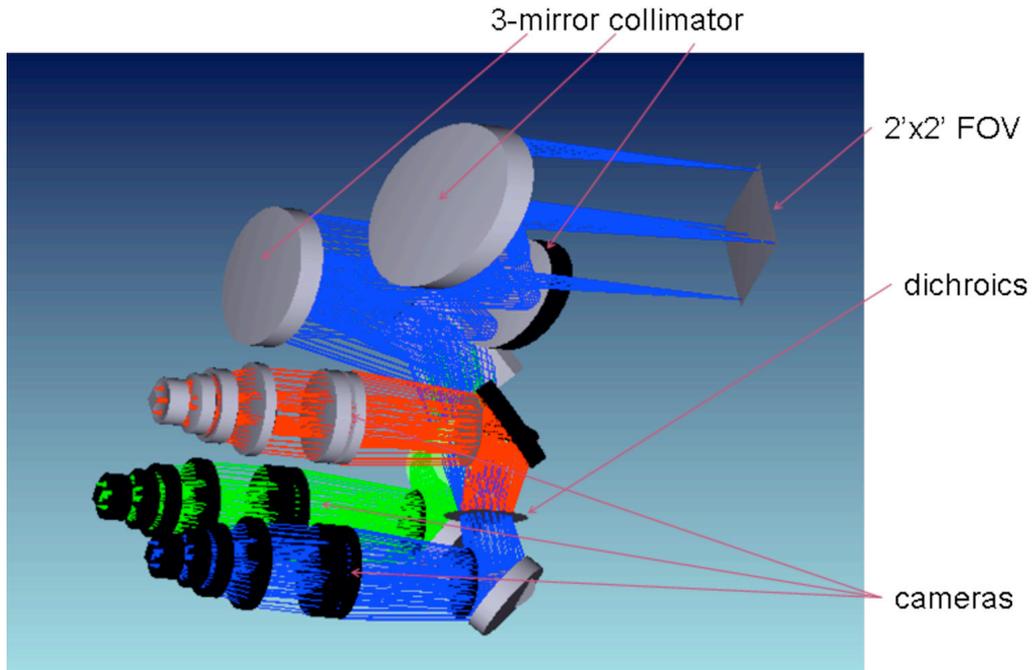

Figure 4 – Optical design of the NIR arms.

### 2.1.4 Image slicer

In order to achieve the science requirement of higher spectral resolution we have designed an image slicer that will divide an area of 5"x1.5" on the focal plane into 5 slices of 5"x0.3" that will form a virtual slit. With these narrower slits we can reach significantly higher resolutions, of ~3,000. The image slicer will cut the field of view into 5 strips, which will be reimaged along a pseudo-slit at the spectrograph entrance. The central slice does not include any dioptre, as the light is directly transmitted to the spectrograph. Only the four lateral sliced fields are reflected towards the spherical mirrors and re-aligned in order to form the virtual slit (see Figure 5).

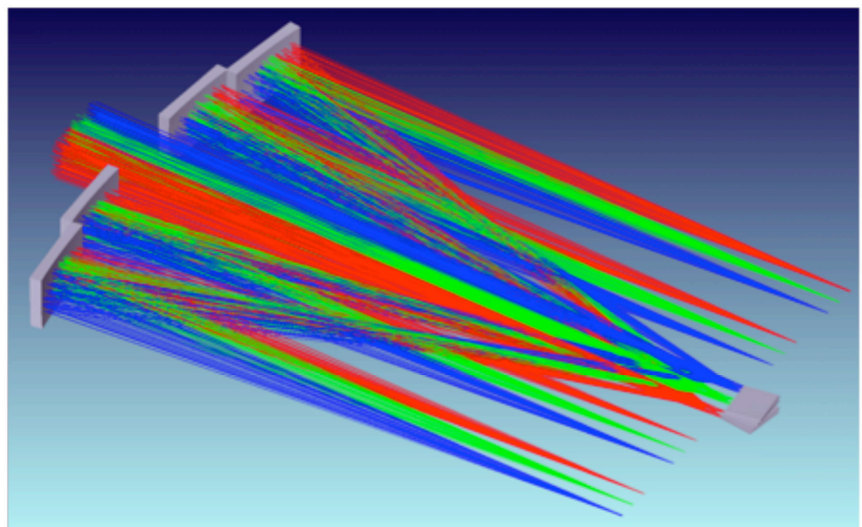

Figure 5 – Design of the image slicer unit.

## 2.2 Mechanical design

The instrument is divided into four main mechanical subassembly units, as shown in Figure 6.

The main structure has been designed and optimized using a welded steel truss. We impose as a requirement that it must be possible to disassemble each arm separately. Low weight and stiffness are the other design drivers.

The UV arm is composed of two subassemblies, UV Box-1 and UV Box-2, mounted directly onto the main structure. This configuration helps with the integration of the whole OCTOCAM before mounting it on to the telescope. UV Box-1 contains the folding mirror 1 and the first lens. UV Box-2 holds the folding mirrors 2 and 3 and the focal reducer. The overall structure is designed to work as an optical bench, able to keep the optical elements within the requirements under all the allowed orientations of the folded Cassegrain station.

The VIS optical elements are mounted onto a single optical bench directly fixed onto the main structure via three referencing spherolinders. The optical elements will be mounted using traditional optomechanical designs, i.e. camera barrels for the lenses and kinematic mountings for the mirrors. Particular effort will be made for the design of the filter/grating wheel in order to cope with the reduced available space. The detector system is designed to reduce weight and volume. There will be two cryostats for the VIS arm. Each cryostat will house two adjacent detectors. Vibrationless helium cryocoolers will be used in order to reduce the weight.

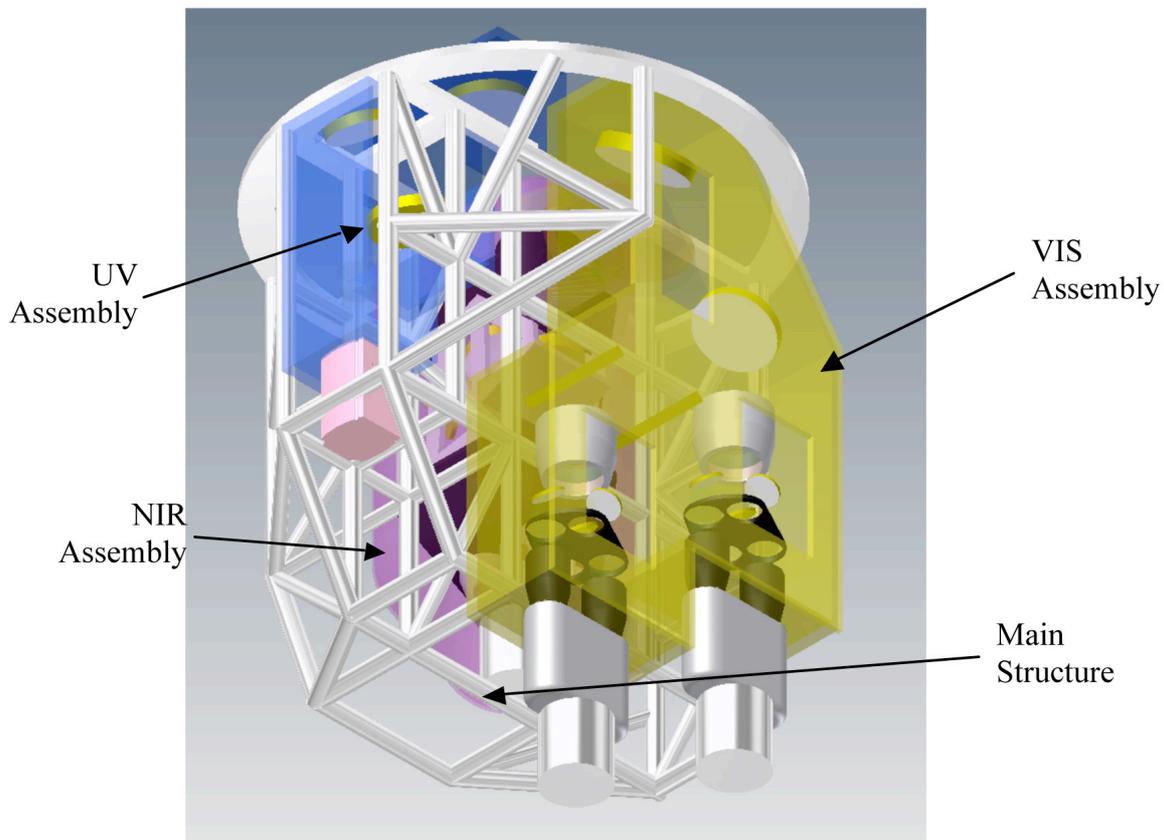

Figure 6 – Mechanical design of OCTOCAM.

The NIR arm has been designed to be directly mounted onto the main structure. The vacuum vessel and the thermal shield hold all the optomechanical elements together. The mechanical layout is composed of two main parallel benches connected by four rods (see Figure 7). The vacuum vessel will have two windows. The main entrance window reflects the UV wavelengths to the UV arm and a secondary window feeds the VIS arm. The focal plane will be populated through a cold slit wheel that positions the slits or the image slicer. The VIS dichroic will be fixed directly to the upper bench. The collimator will be separately mounted and pre-aligned in a dedicated sub-structure. The mirrors will be connected to the upper bench through cryogenic mountings. There will be a cryogenic wheel in order to swap gratings and filters. The focal reducer will be fixed onto the lower bench. The detectors will be fixed onto a separate plate directly linked to the cold finger.

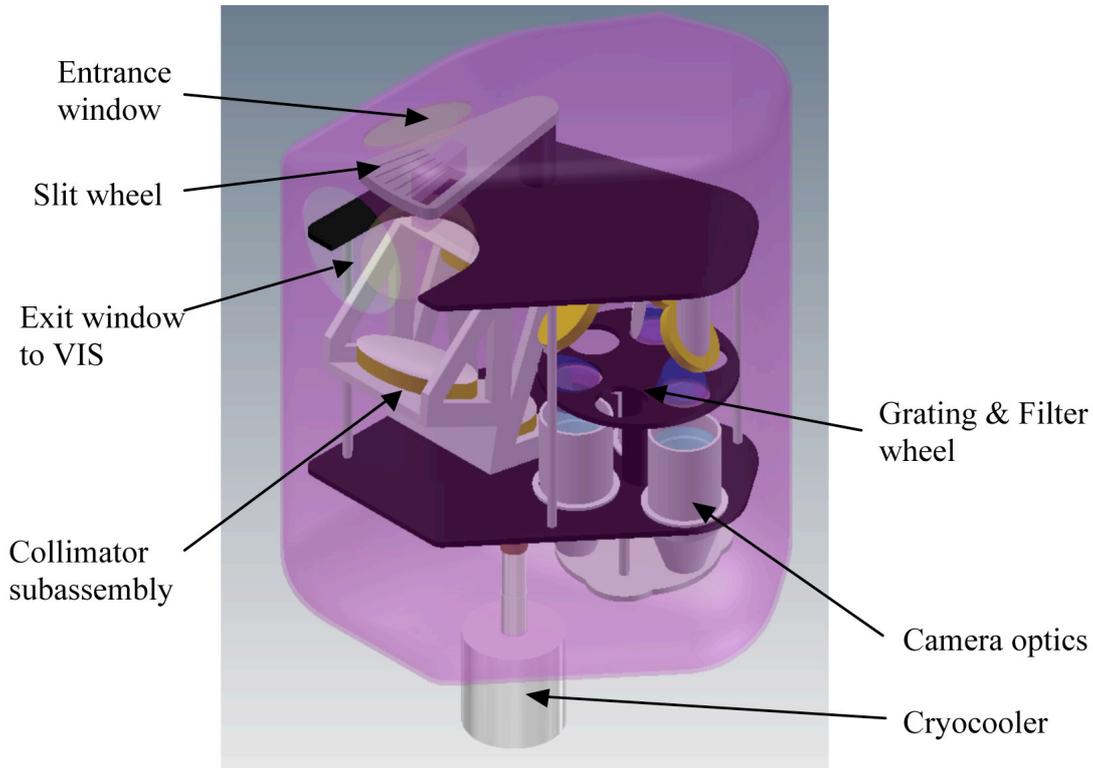

Figure 7 – Cryogenic vessel of OCTOCAM.

## 2.3 Detectors

The eight cameras ($u,g,r,i,z,J,H,K_S$) will belong to only three technology groups (i.e. three types of detectors) - $\{u\},\{g,r,i,z\},\{J,H,K_S\}$ (or UV, VIS and NIR).

The instrument design requires a 1kx1k detector with the highest possible sensitivity in the *u*-band. Our primary choice here is e2v's CCD201-20 EMCCD detector, specifically due to the availability of a commercial camera (Andor iXon+ 888) based on this device [3]. This camera allows 2x2 binning readouts in 57ms, with negligible readout noise thanks to the electron multiplying capability.

For the VIS arms we have the widest range of available options. In this case we require at least 2kx1k detectors in order to perform spectroscopy. Our choice is the 2kx2k CCD231-42 from e2v [4], which we will use as frame transfer device. In its highest speed it will allow us to read the central 1kx1k (in binning 2x2 for imaging) in 27 ms. This implies a readout noise of 11.5 e$^-$, which can go down to 3e$^-$ in the slowest readout mode (reading in 1.32 s) .

The IR technology choice is relatively straightforward as the available choices are much more limited. We plan to use the HAWAII-2RG from Teledyne Scientific Imaging together with the SIDECAR ROIC (Read-Out-Integrated-Circuit). The SIDECAR is a cryogenic readout circuit, effectively a controller-on-a-chip for HxRG detectors, connected to the focal plane array via a flexible ribbon. This represents a huge reduction in mass/power/volume, and the elimination of an entire R & D element from the project. This device will allow a readout speed of 6.5ms with a readout noise of 75 e$^-$, gradually going down to 265 ms and a readout noise of 15 e$^-$.

Figure 9 shows a preliminary diagram with the different exposure times and readout noises expected for each arm.

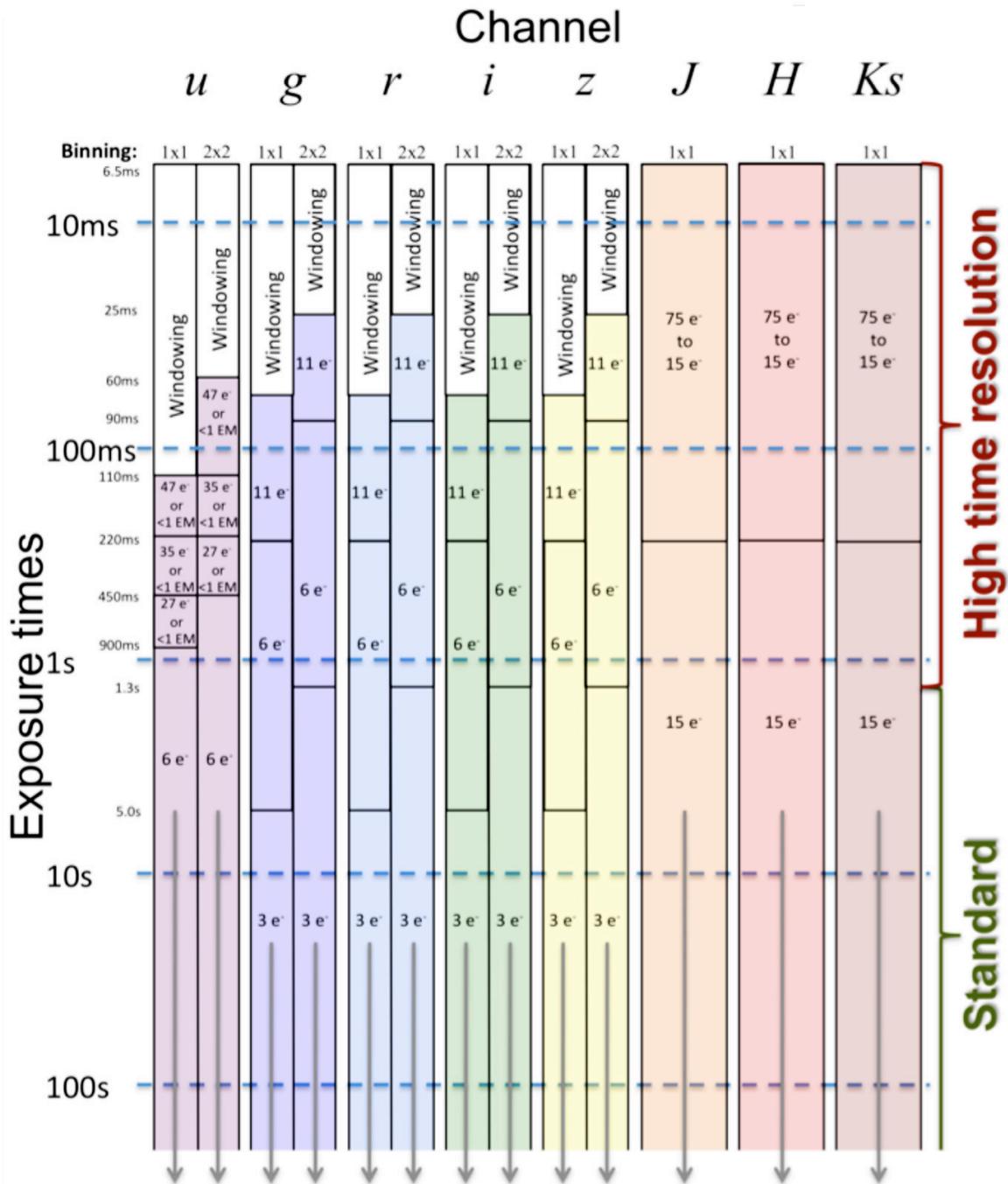

Figure 8 – Exposure time ranges available for each observing channel in imaging mode. For each band the readout noise (in e⁻) is indicated. Note that in *u*-band an EM node will be available, reducing significantly the readout noise. The plot can be divided in two observing regimes; standard slow reading mode (exposure time > 1s) and high-time-resolution mode (exposure time <1s). These two modes can be mixed by the user in different channels.

# 3. PERFORMANCE

A preliminary, conservative efficiency budget has been made considering the transmission efficiency of each of the optical elements in the instrument together with the grisms and detectors. Figure 9 shows the expected efficiencies in both imaging and spectroscopic modes.

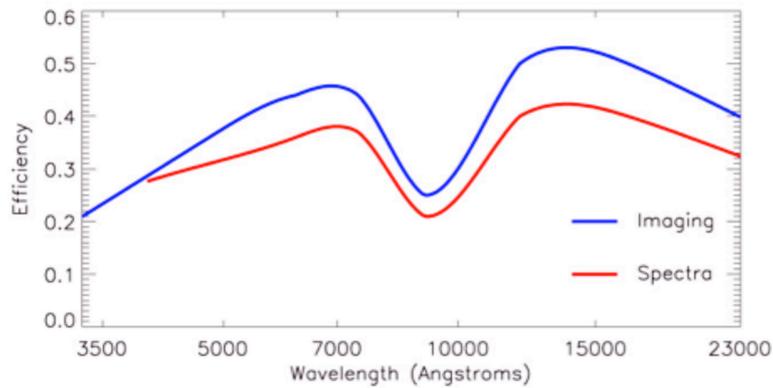

Figure 9 – Efficiency of the instrument.

Using these estimates, together with the telescope mirror reflectivity and the absorption of the atmosphere we have built an instrument performance simulator, with which we have derived the limiting magnitudes shown in the following figure for some representative exposure times.

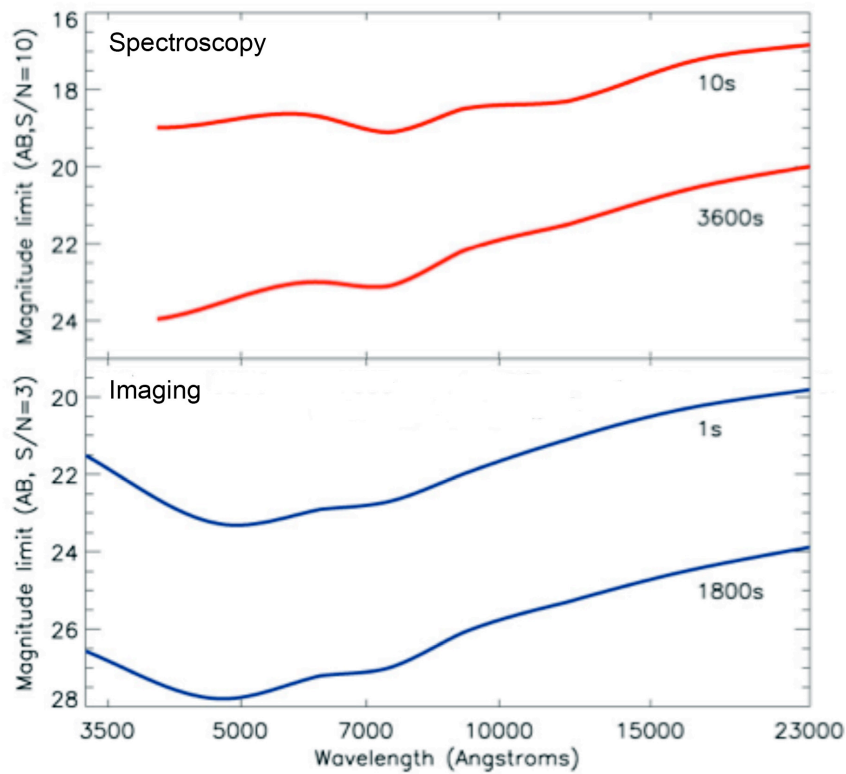

Figure 10 – Spectroscopic (S/N=10) and imaging (S/N=3) limiting magnitudes for OCTOCAM. A power law spectrum $F_\nu \sim \nu^{-1}$, seeing of 0.8", an airmass of 1.2, a slit width of 1.0" (for spectroscopy) and 3 days from the new Moon were assumed.

## 4. SCIENCE CASES

Multiband observations over a wide wavelength range are required for the study of most astrophysical phenomena, making the science cases of OCTOCAM very varied. Having, on top of this, high-time-resolution capabilities opens the door to a further range of scientific possibilities. Here we make a rough summary of some of the science cases that will be possible to undertake with OCTOCAM, from the nearby solar system bodies, to the furthest objects in the Universe.

- *Solar system:* Multiband high-time-resolution can be used to observe stellar occultations by planets and other solar system objects will give us an insight to their morphological properties and composition.

- *Extrasolar planets:* OCTOCAM will be used to observe transits of extrasolar planets. High-timing-accuracy and sensitivity will allow the detection and characterization of planets with both imaging and spectroscopy.

- *Asteroseismology:* The 8-band simultaneous coverage will be used to detect color variation phases as a function of the observed wavelength.

- *Cataclysmic variables:* Through time-resolved, high-sensitivity observations we will be able to derive the mass of the two components, being able to determine the SED of the accretion disk and the characteristics of the secondary star. OCTOCAM will also allow the detailed investigation of the nature of second-scale fluctuations.

- *X-ray binaries:* OCTOCAM will allow the study of jet physics by observing the variability in time scales of 10 -100 ms and to understand the geometry and the equation of state of the neutron stars.

- *Magnetars:* With the simultaneous observation of the optical-NIR bands and its sensitivity it will constrain the spectral energy distribution of faint magnetars. For bright cases we will be able to obtain spectroscopy and time-resolved analysis of flaring episodes.

- *Transients:* The large number of forthcoming surveys will detect many transient sources of varied and sometimes unknown origin. OCTOCAM will be one of the most efficient tools for characterizing those sources thanks to its multirange coverage and sensitivity.

- *Blazars:* High time resolution of the order of one second and sensitivities of mmag will be needed to construct the spectral evolution light curves of flares that are powerful discriminators of the underlying physical models.

- *Supernovae:* Simultaneous coverage in 8 bands will allow follow up of a large number of events with better efficiency and sensitivity than ever before. The NIR channels will also allow the estimation of the fraction of hidden (optically extinguished) star formation rate.

- *Gamma-ray bursts:* Thanks to the simultaneity of the observations with 8 arms and the capacity to do both imaging and spectroscopy, OCTOCAM will be the perfect tool to identify and measure redshifts of the furthest objects in the Universe, allowing to measure them up to redshifts of $z=18$. Furthermore, the high-time-resolution will allow the study of these events which are known to have variability of milliseconds with unparalleled accuracy.

## 5. CONCLUSIONS

OCTOCAM is a workhorse instrument for the GTC, with imaging and spectroscopic capabilities. It will provide a wide-wavelength coverage and high-time-resolution, fulfilling the needs of a very wide range of scientific projects and, with its unique characteristics, open a window to new lines of research. It will cover studies within the Solar System, where it will observe asteroids and comets to allow a better understanding of their mineralogical history. Within the boundaries of our Galaxy, OCTOCAM will be used to detect and characterize exoplanets, study diverse stellar populations or understand the extreme physics of X-ray binaries. Thanks to the large collecting area of GTC and the sensitivity of the instrument, it will be also used to study distant galaxies and gamma-ray bursts, the furthermost objects detectable in the Universe that will teach us about our cosmological origin.

In order to achieve these scientific goals we have conceived an instrument capable of observing simultaneously in 8 bands, from the ultraviolet to the near-infrared to multiply the power of what is already the largest telescope in the world. It will do both imaging and spectroscopy in low and medium spectral-resolution modes. Furthermore, using new detector technologies we go one step further offering these already unique capabilities together with high-time-resolution. In this way we merge together in OCTOCAM the experience of instruments like X-shooter [7], GROND [8] and ULTRACAM [9] to create a versatile tool for astronomical observations.

The design presented here shows that it is feasible to build OCTOCAM, meeting the scientific and technical requirements, while keeping within the mass, torque and volume constraints of the Folded Cassegrain focus of GTC.

## REFERENCES

1. www.gtc.iac.es
2. Álvarez, P. et al., "The GTC project: under commissioning" Proc. SPIE 7012,701202-701202-12 (2008).
3. www.andor.com/pdfs/specs/L888SS.pdf
4. www.e2v.com
5. www.teledyne-si.com/imaging/H2RG.pdf
6. http://www.teledyne-si.com/imaging/SIDECAR ASIC Development Kit Brochure - Public.pdf
7. D'Odorico, S., et al., "X-shooter UV- to K-band intermediate-resolution high-efficiency spectrograph for the VLT: status report at the final design review" Proc. SPIE 6269, 626933 (2006).
8. Greiner, J. et al., "GROND—a 7-Channel Imager" The Publications of the Astronomical Society of the Pacific, 120, 866, 405-424 (2008).
9. Dhilon, V. S. et al. "ULTRACAM: an ultrafast, triple-beam CCD camera for high-speed astrophysics", Monthly Notices of the Royal Astronomical Society, 378, 3, 825-840 (2007).